\documentstyle[aps,epsf,twocolumn]{revtex}

\begin{document}
\twocolumn[
\begin{center}
\vspace{0mm}
{\large{\bf Gap Anisotropy and de Haas-van Alphen Effect in Type-II Superconductors
}}

\vspace{5mm}
{Kouji Yasui
and Takafumi Kita}
\par

\vspace{1mm}
{\small\em 
Division of Physics, Hokkaido University,
Sapporo 060-0810, Japan}
\par

({\today})
\vspace{3mm}

{\small \parbox{142mm}{\hspace*{5mm}
We present a theoretical study on the de Haas-van Alphen (dHvA) oscillation
in the vortex state of type-II superconductors, with a special focus on the connection 
between the gap anisotropy and the oscillation damping.
Numerical calculations for three different gap structures clearly indicate
that the average gap along the extremal orbit is relevant for 
the magnitude of the extra damping,
thereby providing support for experimental efforts 
to probe gap anisotropy through the dHvA signal.
We also derive an analytic formula for the extra damping
which will be useful to estimate
angle- and/or band-dependent gap amplitudes.
}
}
\end{center}

\vspace{5mm}
]

A considerable number of materials have been found in the 1990s
to exhibit de Haas-van Alphen (dHvA)
oscillation in the superconducting vortex state\cite{Janssen98},
the phenomenon discovered by Graebner and Robbins
in 2H-NbSe$_{2}$\cite{GR76}.
Many theories have been put forward during the same period
to explain this fundamental phenomenon observed in the system
without a well-defined Fermi surface\cite{Janssen98}.
However, there is yet no established theory comparable
with the normal-state Lifshitz-Kosevich (LK) theory\cite{LK55}.
Moreover, what is lacking seems to be a clear physical picture of
the mechanism of the extra oscillation damping in the vortex state.
To improve the situation, 
we present here a numerical study combined with an analytical one.

Novel aspects in our study are summarized as follows:

(i) We perform {\em three-dimensional} numerical
calculations for the dHvA oscillations in the vortex state
using the Bogoliubov-de Gennes equations.
To date, numerical studies have been carried out only for the two-dimensional 
$s$-wave model, such as that of Norman {\em et al}.\ (NMA)\cite{NMA95}.
Our results show that dHvA oscillations in three dimensions
are quantitatively different in the vortex state from those in two dimensions.

(ii) To clarify {\em the connection between 
the oscillation attenuation and the gap anisotropy},
we use three different gap structures:
$s$-wave, $d$-wave with four point nodes in the extremal orbit,
and $p$-wave with a line node in the extremal orbit; see Eq.\ (\ref{gap})
below.
It is thereby shown that the attenuation is determined by the
average gap along the extremal orbit in zero magnetic field (see Fig.\ 1),
in disagreement with the theory of Miyake\cite{Miyake93}.
This result still indicates that
one can probe the angle- and/or band-dependent gap amplitude 
in zero field with the dHvA effect in the vortex state.
The origin of the discrepancy from Miyake's
result is discussed in some detail.

(iii) We derive {\em a new analytic formula for the extra Dingle temperature}
in the vortex state:
\begin{eqnarray}
k_{\rm B}T_{\Delta}=
0.5\hspace{0.3mm}\tilde{\Gamma}\hspace{0.3mm}\langle|\Delta_{\bf p}|^{2}\rangle_{\rm eo}
\hspace{0.3mm}\frac{m_{\rm b}c}{\pi e\hbar}
\, \frac{1\!-\!B/H_{c2}}{B} \, .
\label{T_D}
\end{eqnarray}
Here, $\tilde{\Gamma}\!=\!0.125$ is a dimensionless quantity 
characterizing the Landau-level broadening due to the pair potential,
$\langle|\Delta_{\bf p}|^{2}\rangle_{\rm eo}$ denotes the average gap 
along the extremal orbit in zero field,
$m_{\rm b}$ is the band mass,
and $B$ and $H_{c2}$ are the average flux density and the upper critical field,
respectively.

Equation (\ref{T_D}) is derived below through the second-order perturbation
with respect to the pair potential, i.e., an approach from $H_{c2}$, which
is useful to estimate the gap amplitude along the extremal orbit.
A difference from Maki's formula\cite{Maki91} lies in the prefactor
where the Fermi velocity $v_{\rm F}$ is absent.
Indeed, a dimensional analysis on the second-order perturbation
tells us that the Landau-level broadening in the vortex state
should be of order
$|\tilde{\Delta}^{\!(0)\!}(B)|^{2}/\hbar \omega_{\rm c}$,
where $\hbar \omega_{\rm c}$ is the cyclotron energy
and $\tilde{\Delta}^{\!(0)\!}(B)\!\propto\! \sqrt{\langle|\Delta_{\bf p}|^{2}\rangle_{\rm eo}
(1\!-\! B/H_{c2})}$ is essentially the average gap along the extremal orbit.
This leads to Eq.\ (\ref{T_D}) except for the numerical constant.

Terashima {\em et al}.\cite{Terashima97} reported a dHvA experiment on
YNi$_{2}$B$_{2}$C
where an oscillation (labeled $\alpha$) is seen to persist down to a
field of order $0.2H_{c2}$.
On the other hand, a specific-heat experiment at $H\!=\!0$
shows a power-law behavior $\propto\! T^{3}$ 
at low temperatures\cite{Movshovich94},
indicating the existence of the gap anisotropy in this material.
This gap anisotropy may also play an important role for 
the dHvA signal far below $H_{c2}$.
By using Eq.\ (\ref{T_D}), 
the average gap of this band $\alpha$
will be estimated at the end of the paper.

{\bf Model}: Our starting point is the Bogoliubov-de Gennes equation
for the quasiparticle wavefunctions
${\bf u}_{s}$ and ${\bf v}^{*}_{s}$
labeled by a quantum number $s$
with a positive eigenvalue $E_{s}$:
\begin{eqnarray}
&& \int \! d{\bf r}_{2}\left[ 
\begin{array}{cc}
\vspace{1mm}\,\,\, \underline{{\cal H}}({\bf r}_{1},{\bf r}_{2}) & \,\,
\underline{\Delta} ({\bf r}_{1},{\bf r}_{2}) \\ 
-\underline{\Delta}^{\! *}\!({\bf r}_{1},{\bf r}_{2}) &\! -\underline{{\cal H}%
}^{*}\!({\bf r}_{1},{\bf r}_{2})
\end{array}
\right] \left[ 
\begin{array}{c}
\vspace{1mm} \,\,{\bf u}_{s}({\bf r}_{2}) \\ 
-{\bf v}^{*}_{s}({\bf r}_{2})
\end{array}
\right] \hspace{5mm}  \nonumber \\
&& \hspace{25mm}= E_{s}\! \left[ 
\begin{array}{c}
\vspace{1mm} \,\,{\bf u}_{s}({\bf r}_{1}) \\ 
-{\bf v}^{*}_{s}({\bf r}_{1} )
\end{array}
\right] \, .
\label{BdG}
\end{eqnarray}
Here $\underline{\Delta}$ is the pair potential and $\underline{{\cal H}}$
denotes the normal-state Hamiltonian in the external field
${\bf H}\!\parallel\!\hat{\bf z}$; 
both are $2\!\times\!2$ matrices to describe the
spin degrees of freedom.
We adopt as $\underline{{\cal H}}$ the free-particle Hamiltonian.
As for the pair potential, 
we wish to consider three cases to yield
the following energy gaps at $H\!=\!0$:
\begin{eqnarray}
{\underline \Delta}_{{\bf p}}=
\left\{ 
\begin{array}{l}
\vspace{1mm} \Delta_{0} W_{p}\,i {\underline \sigma}_{y} \\ 
\vspace{1mm} \Delta_{0} W_{p}\sin^{2}\!\theta_{\bf p}
\cos 2\varphi_{\bf p}\,i{\underline \sigma}_{y}\\
\Delta_{0} W_{p}\cos\theta_{\bf p} \,
i{\underline \sigma}_{z}{\underline \sigma}_{y}
\end{array}
\right. ,
\label{gap}
\end{eqnarray}
where ${\bf p}$ is the momentum,
$W_{p}$ denotes some cut-off function with $W_{p_{\rm F}}\!=\! 1$
($p_{\rm F}$: Fermi momentum),
$\theta_{\bf p}$ ($\varphi_{\bf p}$) is the polar (azimuthal) angle,
and ${\underline \sigma}_{j}$'s are the Pauli matrices.
The first one is the isotropic $s$-wave state,
whereas the latter two have four point nodes ($d$-wave) and a line node
($p$-wave), respectively, in the extremal orbit of the $xy$ plane perpendicular
to ${\bf H}\!\parallel\!\hat{\bf z}$.
It then turns out 
that the orbital part of the corresponding pair potentials 
in finite fields
can be expanded with respect to
${\bf r}\!\equiv\!{\bf r}_{1}\!-\!{\bf r}_{2}$ and
${\bf R}\!\equiv\!\frac{1}{2}({\bf r}_{1}\!+\!{\bf r}_{2})$ as\cite{Kita98}
\begin{eqnarray}
\Delta({\bf r}_{1},{\bf r}_{2})=&& \frac{{\cal N}_{\rm f}}{\sqrt{2}}
\sum_{N_{\rm c}}\tilde{\Delta}^{(N_{\rm c})}(B)\,
\psi_{N_{\rm c}{\bf q}}^{({\rm c})}({\bf R}_{\perp})
\!\!\sum_{N_{\rm r}m p_{z}}\!
(-1)^{N_{\rm r}}
\nonumber \\
&& \hspace{-2mm}\times\, \psi_{N_{\rm r}m}^{({\rm r})}({\bf r}_{\perp})\, 
\frac{{\rm e}^{ip_{z}z/\hbar}}{L_{z}}
\times
\left\{ 
\begin{array}{l}
\vspace{1mm}  W_{p} \\ 
\vspace{1mm}  \frac{1}{2}W_{p}\sin^{2}\!\theta_{\bf p} \\
              W_{p}\cos\theta_{\bf p}
\end{array}
\right. .
\label{LL}
\end{eqnarray}
Here $N_{\rm c}$ and $N_{\rm r}$ denote the Landau levels
in the average flux density $B$,
${\bf q}$ is an arbitrarily chosen center-of-mass magnetic Bloch vector,
and 
$m$ signifies the relative angular momentum along the $z$
axis so that $m\!=\! 0$, $0$, and $\pm 2$ for the three cases, respectively,
in a system with ${\cal N}_{\rm f}^2/2$ flux quanta and the length $L_{z}$
along the $z$ axis.
The arguments ${\bf r}_{\perp}$ and ${\bf R}_{\perp}$ denote the $xy$ components,
and $p$ and $\theta_{\bf p}$ are to be evaluated at
$p\!=\!(p_{z}^2\!+\!\hbar^{2}\hspace{-0.5mm} N_{\rm r}/l_{B}^{2})^{1/2}$
and
$\theta_{\bf p}\!=\!\tan^{-1}\frac{\hbar\sqrt{N_{\rm r}}/l_{B}}{p_{z}}$
with $l_{B}\!\equiv\!(\hbar c/eB)^{1/2}$.
See ref.\ \onlinecite{Kita98} for the expressions of 
the basis functions $\psi_{N_{\rm c}{\bf q}}^{({\rm c})}$ and 
$\psi_{N_{\rm r}m}^{({\rm r})}$.

A significant advantage of Eq.\ (\ref{LL}) is that the coefficients 
$\{\tilde{\Delta}^{(N_{\rm c})}\}_{N_{\rm c}=0}^{\infty}$ completely specify
the pair potential, and the first few terms
suffice to describe those of $B\! 
\begin{array}{c} > \vspace{-2.7mm}  \\  \sim \end{array} 
\! 0.1H_{c2}$\cite{Kita98-2,Yasui00}.
The self-consistent mean-field theory\cite{NMA95,Rasolt92,Yasui00}
predicts an oscillatory reentrant behavior of $H_{c2}$ and 
$\tilde{\Delta}^{(N_{\rm c})}$.
On the contrary,
such a singular behavior of $H_{c2}$ has never
been identified definitely in any materials displaying the 
dHvA oscillation in the vortex state.
Leaving the puzzling discrepancy for a future study,
we here adopt the quasiclassical
$\tilde{\Delta}^{(N_{\rm c})}$
rather than the fully self-consistent one,
with the square-root behavior
\begin{equation}
\tilde{\Delta}^{\!(0)}= a(1\!-\! B/H_{c2})^{1/2}
\label{Del(0)}
\end{equation}
of the mean-field second-order transition
for the dominant $N_{\rm c}\!=\! 0$ level.
Then the best choice for $\{\tilde{\Delta}^{(N_{\rm c})}\}_{N_{\rm c}=0}^{\infty}$
would be to use the results from
the Eilenberger equations.
Since our main interest lies in studying the differences in 
the oscillation damping among various gap structures, however,
we here adopt a model form of $\tilde{\Delta}^{(N_{\rm c})}$
determined by requiring that the maximum of
$\frac{1}{V}\!\int \! d{\bf R}\, |\! \int \!
d{\bf r}\,\Delta({\bf r}_{1},{\bf r}_{2}) {\rm e}^{-i{\bf p}\cdot{\bf r}/\hbar}|^{2}$
be equal to $\Delta_{0}^{2}(1\!-\! B/H_{c2})$,
where $V$ is the volume of the system
and $\Delta_{0}$ denotes the maximum energy gap of the weak-coupling theory 
at $T\!=\! H\!=\! 0$.
This is possible within the lowest-Landau-level approximation of 
retaining only $\tilde{\Delta}^{\!(0)\!}$,
which is excellent for $B\! 
\begin{array}{c} > \vspace{-2.7mm}  \\  \sim \end{array} 
\! 0.1H_{c2}$\cite{Kita98-2,Yasui00}.
The resulting $\tilde{\Delta}^{\!(0)\!}$ also 
displays the square-root behavior, and
our numerical calculation shows that $a^{2}\!\approx\!0.5\Delta_{0}^{2}$ 
in all the three cases.
We are planning to report on the best choice for $a$
from the Eilenberger equations.

The use of the quasiclassical pair potential (\ref{Del(0)}) has an
advantage:
Self-consistent calculations for the continuum model\cite{NMA95}
necessarily result in a rather small number of Landau 
levels below the Fermi energy $\varepsilon_{\rm F}$,
i.e., $N_{\rm F}\!\sim\! 10$ at $H_{c2}$,
hence failing to meet the condition $N_{\rm F}\!\gg\! 1$
appropriate for real materials.
The present calculation with Eq.\ (\ref{Del(0)})
is free from such a limitation.
We have performed calculations including about $50$ Landau levels 
below $\varepsilon_{\rm F}$ for the extremal orbit at $H_{c2}$.

Since the relevant materials have large Ginzburg-Landau parameter
$\kappa\!\gg\! 1$,
we also neglect the screening in the magnetic field.
Indeed, the effect has been shown to be
irrelevant for the oscillation damping\cite{Janssen98}.
We put $T\!=\! 0$, adopt
$W_{p}\!=\! {\rm e}^{-(\xi_{p}/0.1\varepsilon_{\rm F})^{4}}$
with $\xi_{p}$ the normal-state one-particle 
energy measured from $\varepsilon_{\rm F}$,
and choose the cyclotron energy at $H_{c2}$
as $\hbar\omega_{{\rm c}2}\!=\!k_{\rm B}T_{c}$,
in accordance with
$\hbar\omega_{{\rm c}2}/k_{\rm B}T_{c}\!=\!1\!\sim\!3$
for real materials.

{\bf Numerical Method}: To solve it numerically for the above model, 
we transform Eq.\ (\ref{BdG}) into 
the eigenvalue problem for the expansion coefficients
of ${\bf u}_{s}$ and ${\bf v}_{s}$ in
the quasiparticle basis functions $\{\psi _{N{\bf k}\alpha }\}$,
where ${\bf k}$ is a quasiparticle magnetic Bloch vector
and $\alpha$ $(=\!1,2)$ signifies twofold degeneracy 
of the orbital states\cite{Kita98}.
Then it can be solved separately for each ${\bf k}\alpha $ due
to the translational symmetry of the vortex lattice.
The overlap integral between $\psi _{N_{%
{\rm c}}{\bf q}}^{({\rm c})}({\bf R}_{\perp})\psi _{N_{%
{\rm r}}m}^{({\rm r})}({\bf r}_{\perp})$ and $\psi _{N_{1}{\bf k}_{1}%
\alpha_{1} }({\bf r}_{1\perp})\psi _{N_{2}{\bf k}_{2}\alpha_{2}}({\bf r%
}_{2\perp})$ vanishes unless ${\bf q}\!=\!%
{\bf k}_{1}+{\bf k}_{2}$ and $\alpha_{1}\!=\!\alpha_{2}$
so that the calculations are simplified greatly.
The corresponding eigenstate is labeled
 explicitly by $s\!=\!(\nu {\bf k}\alpha p_{z}  \sigma )$ with 
$\nu $ ($\sigma $) the band (spin) index.

{\bf Numerical results}: 
Figure 1(a) presents the oscillation of the $s$-wave magnetization
as compared with the normal-state one.
We see clearly that the oscillation frequency is unchanged from that of the normal state.
With $\hbar\omega_{\rm c}\!=\!k_{\rm B}T_{c}$ at $H_{c2}$,
the oscillation is observed to persist down to a rather low field of
$H_{c2}/B \!
\begin{array}{c} < \vspace{-2.7mm}  \\  \sim \end{array} 
\! 1.8$, i.e., $B\!
\begin{array}{c} > \vspace{-2.7mm}  \\  \sim \end{array} 
\! 0.55H_{c2}$, which is lower than $0.8H_{c2}$
where $\hbar\omega_{{\rm c}}$ becomes equal to
the spatial average of the energy gap:
$\Delta_{0}(1\!-\! B/H_{c2})^{1/2}$.
This is partly because the gap is smaller within the extremal orbit,
as shown by Brandt {\em et al}.\cite{Brandt67}
The points with error bars in Fig.\ 1(b) comprise the corresponding Dingle plot
for the extra damping factor $R_{\rm s}$ 
obtained by numerical differentiation.
This extra damping at high fields shows the behavior
$\propto \! 1 \! - \! B/H_{c2}$
in the logarithmic scale,
but an irregularity sets in around $0.55H_{c2}$
where the oscillation disappears.
We attribute this irregularity to the effect of the bound-state formation
in the core region.
The lines are the predictions from various theoretical formulas.
The Maki formula\cite{Maki91} repro-

\begin{figure}[t]
\begin{center}
\leavevmode
\epsfxsize=85mm
\epsfbox{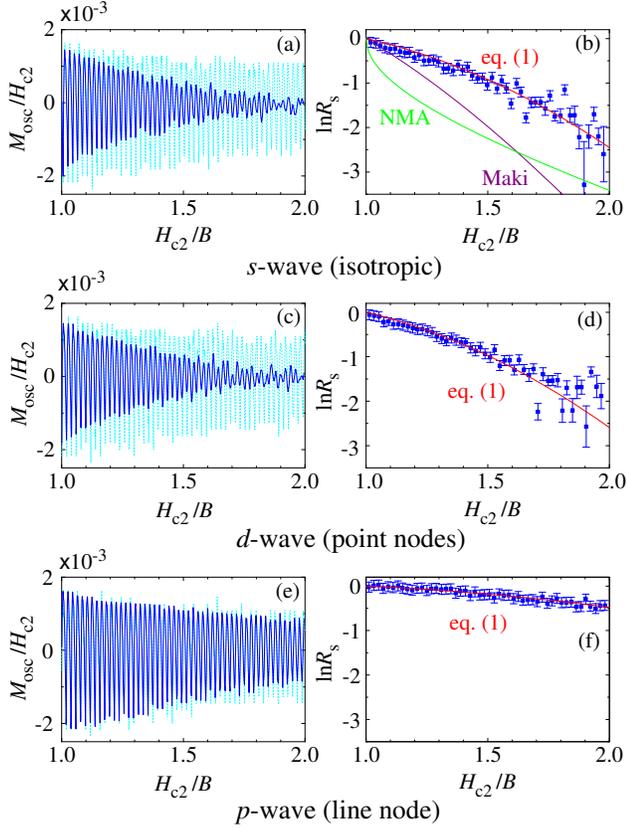}
\end{center}
\caption{Left figures: the oscillatory part $M_{\rm osc}$ 
of the magnetization in the vortex state (full lines)
as compared with the normal-state one (dotted lines)
for the three cases of Eq.\ (\ref{gap}).
Right figures: the corresponding Dingle plots (points with error bars)
compared with theoretical predictions.}
\label{fig:1}
\end{figure}
\noindent
duces the correct functional behavior
$\propto\!1\!-\!B/H_{c2}$ at high fields,
but the prefactor is too large.
The NMA formula\cite{NMA95},
deduced from the two-dimensional self-consistent numerical results 
with $N_{\rm F}\!\sim\!10$ at $H_{c2}$,
clearly shows a more rapid attenuation incompatible with our numerical data.
The theory of Dukan and Te\v sanovi\'c\cite{DT95}, 
which would predict $R_{\rm s}\!=\!0$ in the clean limit of $T\!=\!0$, 
is also inconsistent with the data.
The solid line is due to Eq.\ (\ref{T_D}) derived below; 
it reproduces both the functional dependence 
and the attenuation magnitude of the numerical data.

The above discrepancy between our numerical results and those of NMA
may be attributed mainly to a difference in dimension.
Indeed, the dHvA signal in three dimensions 
differs from that of two dimensions in that some finite region 
$\delta p_{z}$ around the extremal orbit is relevant.
Most of the Landau levels in the region 
do not satisfy the particle-hole
symmetry with respect to $\varepsilon_{\rm F}$, 
so that the effect of the pair potential is rather small
to be handled with the second-order perturbation.
This may be the reason why the oscillation is more persistent 
in our three-dimensional numerical results.
This point is crucial in our derivation 
of the analytic formula below.

Figures 1(c) and 1(d) show the results for the second energy gap in Eq.\ (\ref{gap})
which vanishes at four points on the Fermi surface along the extremal orbit.
The damping is seen to be strong and not much different from the $s$-wave case.
Equation (\ref{T_D}) gives a good fit to the numerical data for
$H_{c2}/B \!
\begin{array}{c} < \vspace{-2.7mm}  \\  \sim \end{array} 
\! 1.8$, thereby indicating that 
the average gap along the extremal orbit is relevant
for the extra damping.
This result is in disagreement with
Miyake's theory that point nodes in the extremal orbit 
should weaken the attenuation\cite{Miyake93}.
Indeed, Miyake obtained his analytic formula by applying a
semiclassical quantization condition 
for the expression of the electron number $N$ at $B\!=\!0$.
Neither of his starting point $N(B\!=\!0)$ nor the use of the quantization
condition may be justified for describing the dHvA signals observed
mainly near $H_{c2}$.

In the line-node case of Figs.\ 1(e) and 1(f),
in contrast, we observe a much weaker damping
in favor of Miyake's idea\cite{Miyake93}.
However, the non-zero extra damping can only be explained
by considering the contribution of some finite width near the extremal orbit.
Using Eq.\ (\ref{T_D}) to obtain the best fit to the numerical data,
i.e., the solid line of Fig.\ 1(f),
we estimate that the region $|p_{z}|\! 
\begin{array}{c} < \vspace{-2.7mm}  \\  \sim \end{array} 
\!p_{\rm F}\sin\!\frac{\pi}{5}$
contributes to the extra attenuation.
This is in rough agreement with the estimation from the Fresnel 
integral $\int_{-\infty}^{\infty}\!
\exp[-i(\sqrt{2\pi N_{\rm F}}\,p_{z}/p_{\rm F})^{2}]\, dp_{z}$
which appears in obtaining the LK formula:
Cutting the infinite integral at 
$\sqrt{2\pi N_{\rm F}}|p_{z}|/p_{\rm F}\!\sim\! 10$ with $N_{\rm F}\!\sim\! 50$
yields a similar value for the relevant range of $p_{z}$.

{\bf Analytic Formula}: The Luttinger-Ward free-energy functional
corresponding to Eq.\ (\ref{BdG}) is given by\cite{Kita96}
\begin{eqnarray}
\hspace{0mm}
\Omega\!=\! -\frac{k_{\rm B}T}{2} \sum_{n} {\rm Tr}\ln \! && \left[ 
\begin{array}{cc}
\vspace{1mm}
\underline{{\cal H}}\!-\!i\varepsilon_{n}\underline{1} & 
\underline{\Delta} \\ 
-\underline{\Delta}^{\! *}& 
-\underline{{\cal H}}^{*}\!-\!i\varepsilon_{n}\underline{1}
\end{array}
\right]
\nonumber \\ 
&& \times\left[ 
\begin{array}{cc}
\vspace{1mm}
{\rm e}^{i\varepsilon_{n}0_{+}}\underline{1}  & \underline{0} \\ 
\underline{0} & 
{\rm e}^{-i\varepsilon_{n}0_{+}}\underline{1}
\end{array}
\right]+{\cdots} \, ,
\label{Omega1}
\end{eqnarray}
where
$\varepsilon_{n}/\hbar$ is the Matsubara frequency
and $0_{+}$ is an infinitesimal positive constant.
The terms $\cdots$ may be expressed solely with respect to 
the pair potential so that they can be neglected in the present model
to consider the oscillatory part.
This $\Omega$ may be transformed into\cite{Kita96}
\begin{eqnarray}
\hspace{0mm}
\Omega\!=\! -\sum_{s}\!\left[ k_{\rm B}T \ln (1\!+\!{\rm e}^{-E_{s}/k_{\rm B}T})
+ E_{s}\!\int\! |{\bf v}_{s}({\bf r})|^{2}\, d{\bf r}\, \right]  \! .
\label{Omega2}
\end{eqnarray}
Since we are interested in the extra damping in the vortex state,
we adopt as $E_{s}$ and ${\bf v}_{s}$ 
the expressions from the second-order perturbation with respect 
to $\underline{\Delta}$. They are
\begin{eqnarray}
E_{N{\bf k}\alpha p_{z}\sigma} \, && = |\xi_{N p_{z}\sigma}|
\! +\! \eta_{N{\bf k}p_{z}}^{(1)}{\rm sign}(\xi_{Np_{z}\sigma}) \, ,
\label{E2nd}
\\
\hspace{0mm}
\int\!\! |{\bf v}_{N{\bf k}\alpha p_{z}\sigma}({\bf r})|^{2} 
d{\bf r} \, &&= \theta(-\xi_{Np_{z}\sigma})
\! +\! \eta_{N{\bf k}p_{z}}^{(2)} {\rm sign}(\xi_{Np_{z}\sigma}) \, ,
\label{vInt}
\end{eqnarray}
where 
$\theta(\xi)$ is the step function
and 
$\eta_{N{\bf k}p_{z}}^{(n)}\!\equiv\!\sum_{N'}\bigl|\int\!\!\int$
$ \psi_{N{\bf k}\alpha\!}^{*}({\bf r}_{1\perp})
\psi_{N'{\bf q}-{\bf k}\hspace{0.2mm}\alpha\!}^{*}({\bf r}_{2\perp})
\frac{{\rm e}^{-ip_{z}(z_{1}-z_{2})/\hbar}}{L_{z}}
\Delta({\bf r}_{1},{\bf r}_{2}) d{\bf r}_{1}d{\bf r}_{2}\bigr|^{2}/$ 
$(\xi_{Np_z\sigma}\!+\!\xi_{N'\! -p_{z}-\sigma})^{n}$.
The first terms on the right-hand side of these equations are just
the normal-state expressions.
The second terms, on the other hand,
denote the finite quasiparticle dispersion 
in the magnetic Brillouin zone
and the smearing of the Fermi surface, respectively,
due to the scattering by the growing pair potential.
It is useful
to express $\eta_{N{\bf k}p_{z}}^{(n)}$ in terms of 
$\tilde{\Delta}^{\!(0)\!}(B)$ and the cyclotron energy $\hbar\omega_{\rm c}$ 
of the extremal orbit as
\begin{eqnarray}
\eta_{N{\bf k}p_{z}}^{(n)} =
\frac{|\tilde{\Delta}^{\!(0)\!}(B)|^{2}}{(\hbar \omega_{\rm c})^{n}} \,
\tilde{\eta}_{N{\bf k}p_{z}}^{(n)}
 \, .
\label{eta}
\end{eqnarray}
The quantity $\tilde{\eta}_{N{\bf k}p_{z}}^{(n)}$ thus defined 
is dimensionless, 
and we realize that 
the main $B$ dependence in Eq.\ (\ref{eta})
lies in the prefactor $|\tilde{\Delta}^{\!(0)\!}(B)|^{2}/(\hbar \omega_{\rm c})^{n}$.
The explicit expression of $\tilde{\eta}_{N{\bf k}p_{z}}^{(n)}$
is given by
\begin{eqnarray}
\tilde{\eta}_{N{\bf k}p_{z}}^{(n)} \! =  &&
\frac{{\cal N}_{\rm f}^{2}}{4}\!\! \sum_{N'mm'}\!\!
\frac{|\langle NN'| 0 \, N\!+\!N'\rangle|^{2}
\langle N\!+\!N'\!+\! m| 2{\bf k}\!-\!{\bf q} \rangle}{
[N\!+\! N'\!-\!2(N_{\rm F}\!+\!\delta)]^{n}} 
\nonumber \\
&& \times
\langle 2{\bf k}\!-\!{\bf q} |  N\!+\!N'\!+\! m' \rangle  \times 
\left\{
\begin{array}{l}
\vspace{1mm} 1 \\
\vspace{1mm} \sin^{4}\!\theta_{\bf p} \\
\cos^{2}\!\theta_{\bf p}
\end{array}
\right. ,
\label{etaT}
\end{eqnarray}
where the overlap integrals are given by eqs.\ (3.23) and (3.29)
of ref.\ \onlinecite{Kita98},
$\delta\!= \!\delta(B,p_{z})$ ($|\delta|\!<\! 1/2$) specifies 
the location of $\varepsilon_{\rm F}$ between the two closest Landau levels,
and $m,m'\!=\! 0$, $0$, and $\pm 2$ 
for the three cases of Eq.\ (\ref{LL}), respectively.
The corresponding normalized density of states:
\begin{eqnarray}
D_{\hspace{-0.2mm}Np_{z}}^{(n)}\hspace{-0.6mm} (\tilde{\eta})
\equiv \frac{2}{{\cal N}_{\rm f}^{2}}\sum_{{\bf k}\alpha}
\delta(\,\tilde{\eta} -\tilde{\eta}_{N{\bf k}p_{z}}^{(n)})\, ,
\label{gDef}
\end{eqnarray}
will play a central role in the following.

Substituting eqs.\ (\ref{E2nd}) and (\ref{vInt}) into Eq.\ (\ref{Omega2}),
we find that the terms containing $\eta_{N{\bf k}p_{z}}^{(2)}$ may be
neglected due to the cancellation between the particle and hole 
contributions.
The remaining term can be transformed with the standard procedure.
We thereby obtain, for the first harmonic of
$\Omega/V$, the expression:
\begin{eqnarray}
\hspace{0mm}
\frac{\Omega_{1}}{V} \,&& =  -\frac{k_{\rm B}T}{2\pi^{2} l_{B}^2} 
\sum_{\sigma}\int_{-1/2}^{\infty} \! \! dN \cos(2\pi N)\!
\int_{-\infty}^{\infty}\!\!dp_{z}\! \int_{-\infty}^{\infty} \!\! d\tilde{\eta} 
\nonumber \\
&&\hspace{0mm}\times 
D_{\hspace{-0.2mm}Np_{z}}^{(1)}\hspace{-0.6mm} (\tilde{\eta}) 
\ln \bigl[1\!+\!{\rm e}^{-(\xi_{Np_{z}\sigma}
+\tilde{\eta}\,|\tilde{\Delta}^{\!(0)\!}(B)|^{2}/\hbar \omega_{\rm c} )/k_{\rm B}T}
\hspace{0.2mm}\bigr]  .
\label{Omega3}
\end{eqnarray}
The function $D_{\hspace{-0.2mm}Np_{z}}^{(1)}\hspace{-0.6mm}(\tilde{\eta})$ 
depends on $(N,p_{z})$,
but may be replaced by a representative one ${\overline D}_{\ell}^{(1)}\!(\tilde{\eta})$
to be placed outside the $N$ and $p_{z}$ integrals\cite{comment1},
where the recovered index $\ell$ specifies the $s$-, $d$-, or $p$-wave case
of Eq.\ (\ref{etaT}).
It may also be acceptable to use a Lorenzian for it:
${\overline D}_{\ell}^{(1)}\!(\tilde{\eta})\!=\!
\tilde{\Gamma}_{\ell}/\pi(\tilde{\eta}^{2}\!+\!\tilde{\Gamma}_{\ell}^{2})$\cite{comment2}.
We thereby obtain an expression for the magnetization
which carries an extra damping factor:
\begin{eqnarray}
R_{\rm s}(B) \, && \equiv \int_{-\infty}^{\infty}\!\!
{\overline D}_{\ell}^{(1)}\!(\tilde{\eta})\exp\!\left[-2\pi i\hspace{0.2mm}
\tilde{\eta}\hspace{0.2mm}
|\tilde{\Delta}^{\!(0)\!}(B)|^{2}/(\hbar \omega_{\rm c})^{2} \hspace{0.2mm}
 \right] d\tilde{\eta} 
\nonumber \\
&& = \exp\bigl[-2\pi\tilde{\Gamma}_{\ell}
|\tilde{\Delta}^{\!(0)\!}(B)|^{2}/(\hbar \omega_{\rm c})^{2} \,\bigr] \, .
\label{R_D}
\end{eqnarray}
Thus, the superconductivity 
gives rise to an extra Dingle temperature of 
$k_{\rm B}T_{\Delta}\equiv \tilde{\Gamma}_{\ell}\,
|\tilde{\Delta}^{\!(0)\!}(B)|^{2}/\pi\hbar \omega_{\rm c}$,
or equivalently,
the extra scattering rate of $\tau_{\rm s}^{-1}\!\equiv\!
2\pi k_{\rm B}T_{\Delta}/\hbar$.

Equation (\ref{R_D}) may have an advantage that one
can trace the origin of the extra dHvA damping
definitely to the growing pair potential,
which brings about a finite quasiparticle dispersion (\ref{etaT})
in the magnetic Brillouin zone
and the corresponding Landau-level broadening, Eq.\ (\ref{gDef}).
Moreover, Eq.\ (\ref{etaT}) reveals that this broadening near $H_{c2}$ is 
closely connected with the zero-field gap structure, Eq.\ (\ref{gap}).

There seems to be no analytic way to estimate $\tilde{\Gamma}_{s}$, so
we determine it using the best fit to the numerical data of Fig.\ 1(b).
Using Eq.\ (\ref{Del(0)}) with $a^{2}\!=\!0.5\Delta_{0}^{2}$,
the procedure yields $\tilde{\Gamma}\!\equiv\!
\tilde{\Gamma}_{s}\!=\!0.125$, as noted before.
It is also clear for the anisotropic cases that 
the average gap around the extremal orbit 
is relevant for the extra attenuation,
as may be realized from Eq.\ (\ref{etaT}).
We hence put $a^{2}\tilde{\Gamma}_{\ell}\!=\!0.5
\langle|\Delta_{\bf p}|^{2}\rangle_{\rm eo}\tilde{\Gamma}_{s}$.
We thereby obtain Eq.\ (\ref{T_D}), 
which gives a good fit to the $d$-wave 
numerical data without any adjustable parameters;
see Fig.\ 1(d).

{\bf Concluding Remarks}: 
We have shown explicitly that the dHvA effect
in the vortex state can be a powerful tool to probe
the average gap along the extremal orbit.
Such an experiment has recently been performed on UPd$_{2}$Al$_{3}$
by Inada {\em et al.}\cite{Inada99}, and
Eq.\ (\ref{T_D}) will be useful in similar experiments
to estimate angle- and/or band-dependent gap amplitudes.
Using the $\alpha$ oscillation of
YNi$_{2}$B$_{2}$C observed by Terashima {\em et al}.\cite{Terashima97},
for example, we obtain $\langle|\Delta_{\bf p}|^{2}\rangle_{\rm eo}\!=\!
1.5$meV for this $\alpha$ band, a value much smaller than 2.5meV
from the specific-heat measurement\cite{Movshovich94}.
It should be noted, however, that the numerical factor $0.5$ 
in Eq.\ (\ref{T_D}) comes from our model described
below Eq.\ (\ref{Del(0)}). 
It has to be replaced by the result from the Eilenberger equations.
We are planning to report on this together with detailed comparisons
with experiments in the near future.

We acknowledge stimulating discussions with Y. Inada
and Z.\ Te\v sanovi\'c.
We are also grateful for the hospitality of
the members of Institut f\"ur Theorie der Kondensierten Materie
at Universit\"at Karlsruhe and 
Physikalisches Institut at Universit\"at Bayreuth,
where a part of the work has been performed.
T.K.\ acknowledges Yamada Science Foundation for financial support.
Numerical calculations were performed on an Origin 2000 in ``Hierarchical
matter analyzing system" at the Division of Physics, Graduate School of
Science, Hokkaido University.



\begin{references}
\bibitem{comment1}The $p_{z}$ integral in Eq.\ (\ref{Omega3})
passes through regions where $\delta\!\approx\!0$ in Eq.\ (\ref{etaT}), 
i.e., the second-order perturbation is not justified.
However, the main contribution to the integral certainly comes from the regions
where we can use Eq.\ (\ref{E2nd}). 
Notice that $\delta\!=\! -1/8$ at the maximal orbit 
when the magnetization takes a local maximum.
\bibitem{comment2}It is found numerically that the variance of Eq.\ (\ref{gDef}) 
depends little on the values of $N_{\rm F}$ and $N$ $(\sim\! N_{\rm F})$,
as expected.
It is $\sim\!0.2$ for the $s$-wave pairing with $\delta\!=\!-1/8$.
\end{references}
\end{document}